\documentclass[prl, twocolumn,preprintnumbers,amsmath,amssymb,english]{revtex4}
\usepackage{dcolumn}
\usepackage{lipsum, babel}
\usepackage{bm}
\usepackage{amsmath,amsfonts,amssymb,ulem}
\usepackage{epstopdf}
\usepackage{xcolor}
\usepackage{color}
\usepackage{wasysym}
\usepackage{graphicx}
\newcommand{\re}[1]{{\color{black}#1}}

\begin{document}

\title{Forward sliding-swing acceleration: electron acceleration by high-intensity lasers\\in strong plasma magnetic fields}

\author{Z. Gong$^{1,2}$}
\author{F. Mackenroth$^{3,4}$}
\author{T. Wang$^{4}$}
\author{X. Q. Yan$^{1}$}
\author{T.  Toncian$^{5}$}
\author{A. V. Arefiev$^{4}$}
\affiliation{$^1$SKLNPT, KLHEDP and CAPT, School of Physics, Peking University, Beijing 100871, China}
\affiliation{$^2$Center for High Energy Density Science, The University of Texas at Austin, Austin, TX 78712, USA}
\affiliation{$^3$Max Planck Institute for the Physics of Complex Systems, 01187 Dresden, Germany}
\affiliation{$^4$University of California at San Diego, La Jolla, CA 92093, USA}
\affiliation{$^5$Institute for Radiation Physics, Helmholtz-Zentrum Dresden-Rossendorf e.V., 01328 Dresden, Germany}

\date{\today}
\begin{abstract}
A high-intensity laser beam propagating through a dense plasma drives a strong current that robustly sustains a strong quasi-static Mega Tesla-level azimuthal magnetic field. The transverse laser field efficiently accelerates electrons in such a field that confines the transverse motion and deflects the electrons in the forward direction, establishing the novel \textit{forward-sliding swing acceleration} mechanism. Its advantage is a threshold rather than resonant behavior, accelerating electrons to high energies for sufficiently strong laser-driven currents. We study the electrons' dynamics by a simplified model analytically, specifically deriving simple relations between the current, the particles' initial transverse momenta and the laser's field strength classifying the energy gain. We confirm the model's predictions by numerical simulations, indicating Mega ampere-level threshold currents and energy gains two orders of magnitude higher than achievable without the magnetic field.
\end{abstract}



\maketitle
Recent advancements in high-power laser technology~\cite{1985_CPA,Mourou_etal_2006} have paved the way for multidisciplinary applications by enabling compact plasma-based sources of energetic particles, such as electrons~\cite{esarey_LWFA_RMP}, ions~\cite{Daido_review_ion,macchi_2013_RMP,Mackenroth_etal_2016}, positrons~\cite{chen2009relativistic_positron,chen2010relativistic_positron}, and neutrons~\cite{pomerantz2014_ultrashort_Neutron}, and radiation~\cite{corde_2013_RMP,AGRThomas_radiation_review}. For these applications the energy transfer from the laser pulse to the plasma electrons is critically important, as once accelerated, they can drive secondary particle and radiation sources. The applications that prioritize the mono-energetic feature of the electron spectrum tend to rely on the laser-wakefield acceleration~\cite{tajima1979,esarey_LWFA_RMP}, whereas the applications that prioritize the electron charge tend to rely on the direct laser acceleration \re{(DLA)} regime~\cite{pukhov1999_DLA,arefiev2012_PRL}. The latter include bright, short-pulsed gamma-ray sources~\cite{nakamura2012high,ridgers2012dense,brady2012laser,ji2014_PRL,ji2014energy,Stark2016_PRL,huang2016_pre,gonoskov2017_prx,zgong_PPCF_2018} that are necessary for advanced nuclear and radiological detection systems~\cite{schreiber2000first_nuclear,kwan2011discrete_nuclear}.



The essence of DLA is an energy transfer from the laser electric field directly to the electrons. This can take place in a dense plasma without stringent density limitations~\cite{liu2013_PRL,Willingale_2018NJP}, which allows the laser to accelerate a large electron population. The regime can even be used to accelerate electrons in optically opaque plasmas if the laser is sufficiently intense to induce relativistic transparency~\cite{2012_Relativistic_transparency,stark2015relativistic,fernandez2017laser}. Typically, the accelerated electrons are pulled into the laser beam from the surrounding plasma with an initially transverse momentum. An electron with initial momentum $p_\text{i}\gg m_e c$ can gain an energy $\varepsilon_0 = \gamma_0 m_e c^2$, where $\gamma _0 \approx (a_0^2 /2) (m_e c/p_\text{i})$, from a plane wave with  intensity $I_0$ and normalized amplitude $a_0 \approx 0.85  I_0 [10^{18} \mbox{ W/cm}^{2}]^{1/2} \lambda[\mu m]$ where $\lambda$ is the wavelength, $m_e$ the electron mass (charge $e<0$), and $c$ the speed of light. The suppression $\gamma _0 \propto 1/p_\text{i}$ is due to the electron dephasing from the laser pulse~\cite{robinson2015_vph,arefiev2016_POP}.




A lot of research has been dedicated to mitigating the negative impact of the dephasing in order to increase the electron energy gain~\cite{pukhov1999_DLA,robinson2013generating,arefiev2015_JPP}. \re{Quasi-static electric fields caused by charge separation have been shown to alter the dephasing, which leads to an enhanced energy exchange between the electrons and the laser~\cite{khudik2016_POP}.} However, this principle is not applicable at next generation laser facilities, such as  ELI~\cite{ELI_laser}, Apollon~\citep{Apllon_laser}, and XCELS~\cite{XCELS}, whose pulses would induce such strong transverse plasma electric fields that the ion motion would become important on time scales shorter than the pulse duration~\cite{wang2019impact}. The resulting transverse ion motion has been shown to dramatically reduce the quasi-static electric fields needed for achieving the electron energy enhancement~\cite{Oliver_2018_PPCF}.

\begin{figure}[t]
\includegraphics[width=0.9\columnwidth]{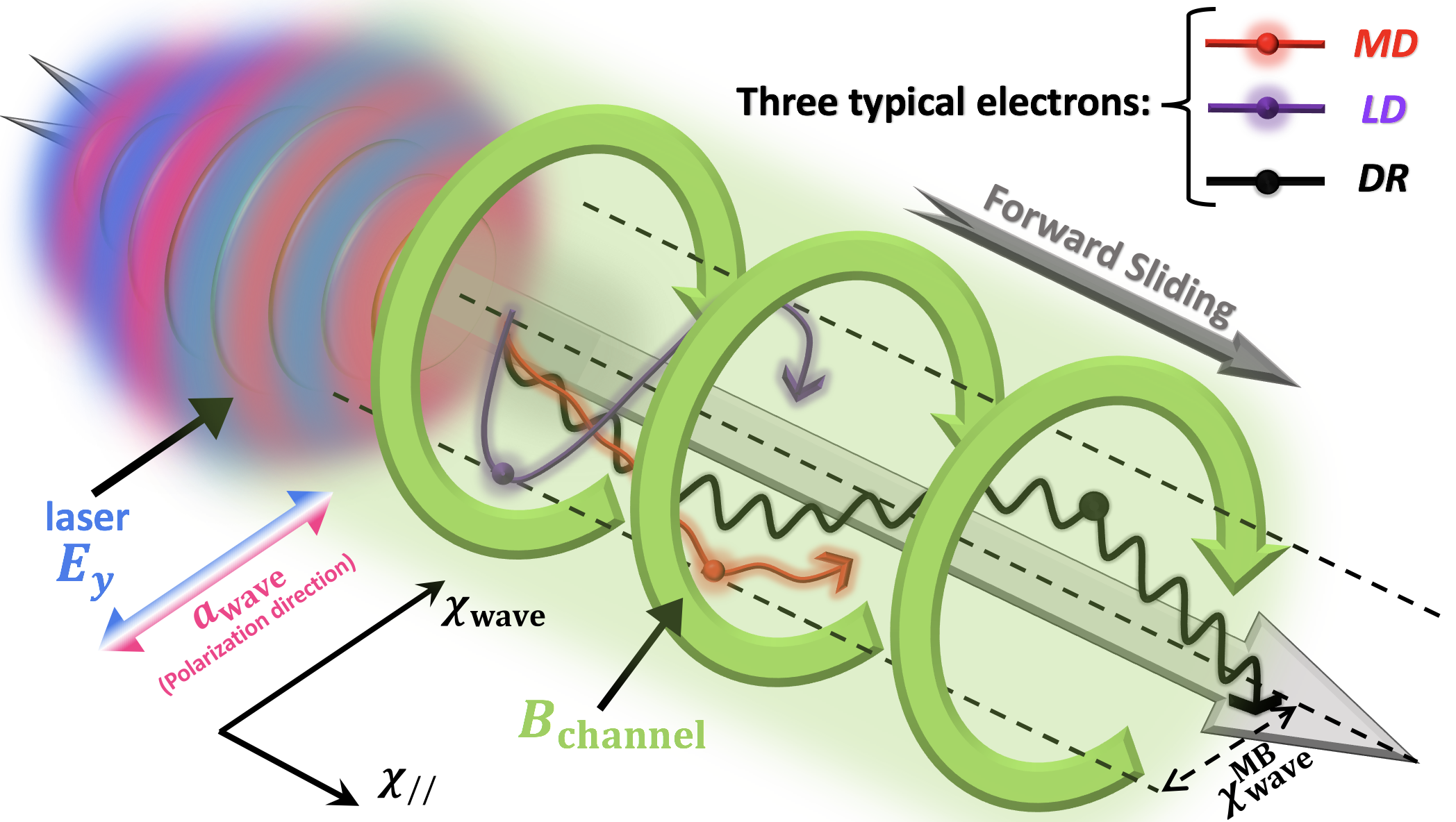}
\caption{Schematic diagram, where the laser propagation along $\chi_\|$ defines the forward sliding direction. The laser polarization is along the direction of $\chi_\text{wave}$. Three different regimes are exemplified: Red, grape and black lines depict momentum dominated (MD),  laser dominated (LD) and deflection regime (DR), respectively.}
\label{fig1}
\end{figure}

Intense lasers additionally drive longitudinal electron currents through the plasma, causing strong quasi-static azimuthal magnetic fields~\cite{lasinski1999particle,bulanov2005ion_MVA,nakamura2010_MVA_PRL,bulanov2010generation_MVA_POP,Oliver_2018_PPCF}. In contrast to the electric fields, these fields are robust with respect to the ion motion and can be sustained at ultra high intensities over hundreds of fs in structured targets~\cite{Supplemental_Material}. \re{Furthermore, in such magnetic fields an electron's motion exhibits a forward drift at constant velocity and dominantly transverse oscillations in a reference frame moving with this drift velocity. This motion can be visualized as a forward sliding swing. In the presence of an additional laser field, the forward sliding motion leads to dephasing such that the laser can directly accelerate the electron along their instantaneous velocity, implying energy gain (s. Fig. \ref{fig1}). Hence, we label the mechanism \textit{forward-sliding swing acceleration} (FSSA).}

In this Letter, we show that \re{in the FSSA regime a} strong azimuthal magnetic field dramatically enhances the electron energy gain from a laser field in a threshold process. We analytically examine the electron dynamics to find a critical current needed to sustain a sufficiently strong magnetic field that leads to an enhanced energy gain regardless of the transverse electron momentum. The advantage of this regime is that it can be employed to generate large numbers of high-energy electrons in an overdense plasma irradiated by an ultra-intense laser (e.g. 2.2 nC above 400 MeV~\cite{Supplemental_Material}). Such dense energetic bunches are the key to driving the bright gamma-ray sources mentioned above.

The model for the FSSA mechanism is based on 3D and 2D particle-in-cell simulations presented in the Supplemental Material~\cite{Supplemental_Material}. The simulations show that (1) a strong current with $J_0 \gg J_A$ can be driven by the laser, where $J_A = m_e c^3 / |e|\approx 17$~kA is the Alfv\'{e}n current~\cite{alfven1939motion}; (2) radial plasma electric fields are much weaker than the azimuthal magnetic field sustained by the current; (3) the laser beam propagation is stable and it is well-approximated by a plane wave with $v_{ph} \approx c$ due to the relativistically induced transparency, where $v_{ph}$ is the phase velocity; (4) the trajectories of energetic electrons remain flat; (5) the magnetic field confines these electrons radially well inside the laser beam; (6) the current density is nearly constant in the region sampled by energetic electrons. We can then capture the characteristic features of the electron dynamics in a single particle model (Fig.~\ref{fig1}) inside a prescribed combination of a plane electromagnetic wave and a static azimuthal magnetic field generated by a homogeneous current $J_0$. The current is characterized by the parameter $\alpha \equiv (\lambda/r)^2 J_0/(4\pi^2 J_A)$.

We consequently study \re{an electron in} a monochromatic, linearly polarized plane laser wave propagating in $x_\|$-direction and polarized along $\bm{e}_\text{wave}$ with electric and magnetic fields given by ${\bm{E}}_\text{wave} = -m_ec/|e|\ \partial{\bm{a}}_\text{wave}/\partial t$ and ${\bm{B}}_\text{wave} = m_ec^2/|e|\nabla \times {\bm{a}}_\text{wave}$ \re{for a vector potential ${\bm{a}}_\text{wave}(t,\bm{x}) = a_0 \cos(\xi+\xi_0) \bm{e}_\text{wave}$, where $\xi=\tau-\chi_\|$ with the dimensionless parameters $\tau=\omega_0t$, $\bm{\chi} = \omega_0\bm{x}/c$, $\bm{\pi}=\bm{p}/mc$, $\gamma=\varepsilon/mc^2$, \re{ where $\omega_0 = (2\pi c)/\lambda$ is the laser frequency, $\varepsilon$ ($\bm{p}$) the electron's instantaneous energy (momentum),} and $\xi_0$ the initial phase.} The azimuthal magnetic field is given by $\displaystyle {\bm{B}}_\text{mag} = m_ec^2/|e|\nabla \times {\bm{a}}_\text{mag} \propto {\bm{e}}_\theta$, derived from ${\bm{a}}_\text{mag} = \alpha (\chi_\text{wave}^2+\chi_\perp^2) {\bm{e}}_\|$ with \re{the reduced coordinate $\chi_\perp$ perpendicular to the laser's propagation and polarization directions}.
The particle dynamics are governed by the relativistic Lorentz equation, which gives the constant of motion $\gamma  - \pi_\| + \alpha \chi_\text{wave}^2 = C_1$, provided radiation reaction \cite{landau1971classical,thomas2012_prx,gonoskov2014anomalous,Cole_etal_2018,Poder_etal_2018} and QED effects \cite{Mackenroth_DiPiazza_2011,di2012extremely,Mackenroth_DiPiazza_2013}, important at higher energies \cite{bulanov2011design,Mackenroth_etal_2013,RRD_bulanov2015}, are negligible. We are going to consider a relativistic electron moving initially in transverse direction \re{$\bm{\pi}(\xi=0)=\pi_\text{i}\bm{e}_{\text{wave}}$} on axis $\chi_{\text{wave}}(\xi=0) = 0$, \re{mimicking the electron injection observed in kinetic simulations~\cite{Stark2016_PRL}}. \re{For $\pi_\text{i}\gg1$} it is $C_1 = \gamma \approx \pi_\text{i}$. We see that \re{without the magnetic field the dephasing $R:= \gamma (d\xi/dt) = \gamma - \pi_\| = \text{const.}$} With the azimuthal magnetic field, however, the dephasing depends on the electron's transverse position according to 
\begin{eqnarray}
R=  \pi_\text{i}-\alpha \chi_\text{wave}^2(\tau,\bm{\chi}). \label{dephasing}
\end{eqnarray}
The transverse momentum $\pi_\text{wave}$ is determined by $d\left( \pi_\text{wave} - a_\text{wave} \right)/d\xi = - (\pi_\|/R) (\partial a_\text{mag}/\partial \chi_\text{wave})$. Substituting $\pi_\|(\chi_\text{wave},d\chi_\text{wave}/d\xi)$ and $R(\chi_\text{wave})$ into the above relation, it can be reduced to the second order nonlinear differential equation 
\begin{eqnarray}
\chi_\text{wave}^{\prime\prime} - \frac{\alpha \chi_\text{wave} \chi_\text{wave}^{\prime2}}{R} + \frac{\alpha \chi_\text{wave}}{R^3} - \frac{\alpha \chi_\text{wave}}{R} = \frac{a_\text{wave}^\prime}{R}, \label{exactdynamics}
\end{eqnarray}
where the prime denotes $d/d\xi$. 

\re{Before turning to a quantitative }numerical solution of Eq.(\ref{exactdynamics}), we provide a qualitative analysis in limiting cases. First the electron's energy $\gamma(\xi)$ evolves according to
\begin{eqnarray}
\frac{d\gamma}{d\xi} = -\left(\frac{\pi_\text{wave}}{R}\right)a_0\sin(\xi+\xi_0).\label{dE/dx}
\end{eqnarray}
%
From this expression we infer that the \re{electron gains energy only from the laser field, as it must be, and that the} laser's periodic oscillations will \re{cancel} any energy gain, unless $R\to0$ over a phase interval $\xi<2\pi$. On the other hand, from Eq.~(\ref{dephasing}) we conclude that $R\to0$ only if $\chi_\text{wave}\to\chi_\text{wave}^\text{MB}:=\sqrt{\pi_\text{i}/\alpha}$. Interestingly, in the geometry introduced above already the azimuthal magnetic field alone will confine the electron motion to transverse excursions $\chi_\text{wave} \leq \chi_\text{wave}^\text{MB}$, whence we label this value the \textit{magnetic boundary} (MB). Since the electron's motion is in $(\chi_\|,\chi_\text{wave})$-plane, the maximum magnetic field, which the electron can experience, is $B_{max}=2\sqrt{\pi_\text{i}\alpha}$ when $\chi_\text{wave} \to \chi_\text{wave}^\text{MB}$. Now we study two particular cases in which $\chi_\text{wave} \to \chi_\text{wave}^\text{MB}$ is possible: the \textit{momentum dominated} (MD) regime ($\pi_\text{i}\gg a_0 \gg1$) and the \textit{laser dominated} (LD) regime ($1<\pi_\text{i}\ll a_0$).
\begin{figure*}[t]
\includegraphics[width=1.9\columnwidth]{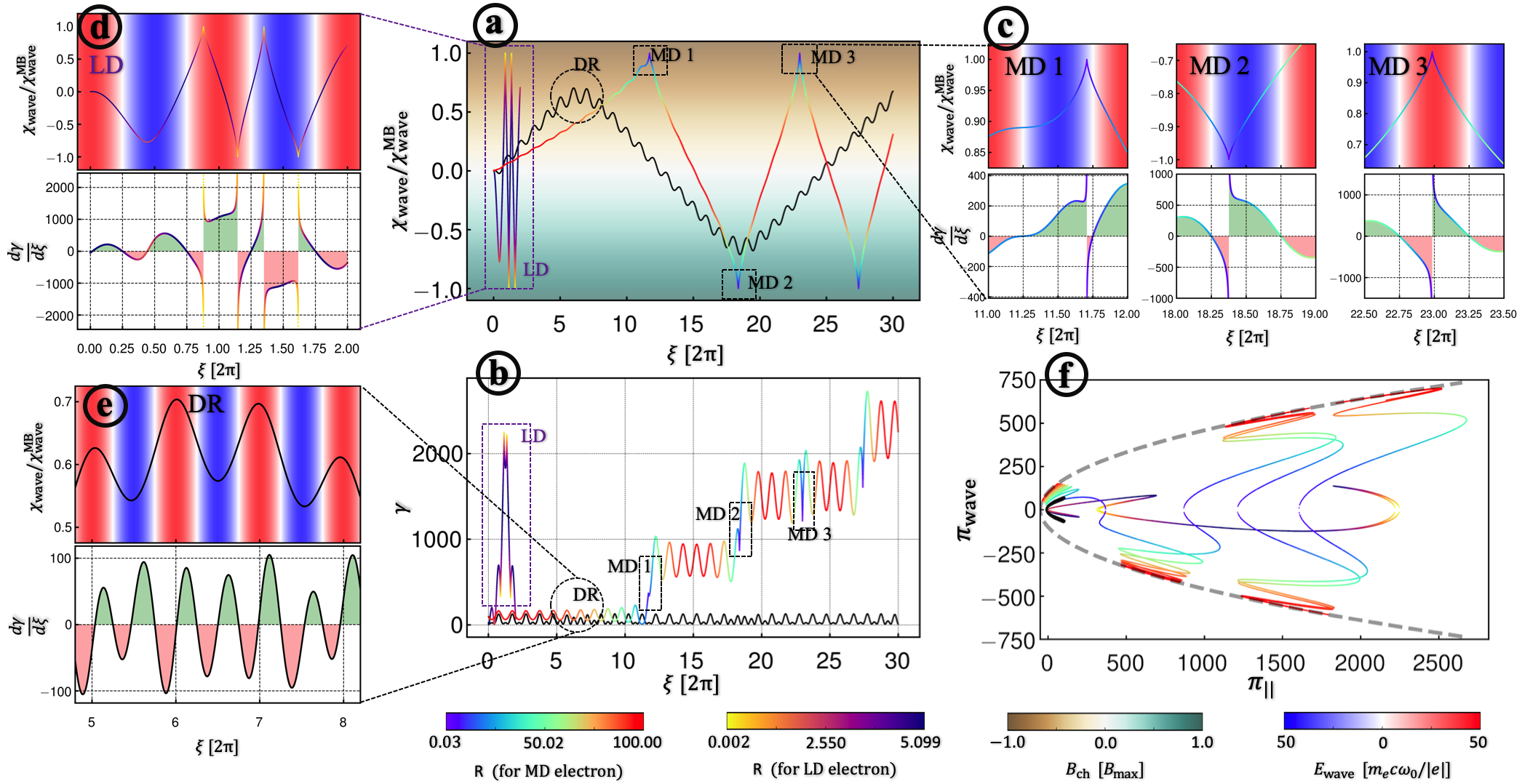}
\caption{(a) Numerically obtained trajectories in three different regimes: MD ($\pi_\text{i}=100$), LD ($\pi_\text{i}=5$) and DR ($\pi_\text{i}=25$) with $R$ in rainbow, magma and black color, respectively. (b) Electron energy (same color code as (a)). (c) Zoom in on MD regime turning points in position (upper panel) and energy space (lower panel) indicating the asymmetry of $d\gamma/d\xi$. (d) and (e) are the same as (c) but for the LD and DR case, respectively. (f) Electron evolution in $(\pi_\|,\pi_\text{wave})$-space, where the dashed gray line indicates the limiting trajectory $\pi_\|=(1+\pi_\text{wave}^2-R^2)/(2R)$ for $R=100.0$.}
\label{fig2}
\end{figure*}

\textit{Momentum dominated regime} --- The laser-induced transverse momentum can be regarded as a small perturbation of the electron's large initial momentum. As discussed above, \re{the magnetic field alone would reflect} the electron at the MB and a comparatively weak laser will not disturb this behavior strongly. We can derive a quantitative threshold above which transverse momenta the MD region is accessed by comparing the time scales on which the magnetic and  laser fields act on the electron, respectively. Obviously the magnetic field will deflect the electron's initial transverse momentum fully into longitudinal motion $\pi^\| \sim \pi^0$ on a time scale $\tau^\text{mag} \sim \chi_\text{wave}^\text{MB}$. \re{On the other hand, averaging the momentum of an electron inside a plane wave \cite{Meyer_1971} over one laser period, it can be estimated that each period the laser field imparts a longitudinal momentum of order $\left<\pi^\|_\text{las}\right>\lesssim a_0^2/\pi_\text{i}$ onto the electron. Hence, in order to impart a longitudinal momentum $\pi^\| \sim \pi_\text{i}$ onto the electron the laser needs a time $\tau^\text{laser} \gtrsim \pi_\text{i}^2/a_0^2$.}
In the MD regime, the laser field must not act on time scales shorter than the magnetic field, or else it would yield the main effect on the electron motion. Consequently, we expect the MD regime to be accessible in the regime $\tau^\text{mag} \leq \rho \tau^\text{laser}$, where $\rho$ is a free parameter accounting for enhancements of the laser acceleration \re{over the case without magnetic field}. Expressed in terms of the initial transverse momentum this condition reads
\begin{eqnarray}
\pi_\text{i} \geq \pi_\text{MD} := a_0^{\frac43} \alpha^{-\frac13} \rho^{-\frac23}.\label{p_MD}
\end{eqnarray}

\textit{Laser dominated regime} --- The initial momentum is small enough for the electron's transverse momentum to be dominated by the laser field. We note that in this regime $\chi_\text{wave}^\text{MB}$ can become small, whereas the plane laser wave transversely displaces the electron by $\chi_\text{p.w.}^\text{max} \sim 1+a_0/\pi_\text{i}$ within one period \cite{Meyer_1971, Arefiev_POP_2014}. As soon as $\chi_\text{p.w.}^\text{max} > \chi_\text{wave}^\text{MB}$ the dominating laser dynamics will drive the electron close to the MB, whence it can again acquire net energy. Neglecting the unity in $\chi_\text{p.w.}^\text{max}$, originating from the electron's initial momentum, we can recast the above condition for the onset of the LD regime to read $ a_0/\pi_\text{i}\geq \kappa \sqrt{\pi_\text{i}/\alpha}$ where $\kappa\in [0,1]$ is a free parameter accounting for deviations from the laser-driven plane wave dynamics inside the weak azimuthal magnetic field. Solving the above condition for $\pi_\text{i}$ we find the LD regime to be accessible for initial electron momenta satisfying
\begin{eqnarray}
\pi_\text{i}\leq \pi_\text{LD} = \alpha^{\frac13} a_0^{\frac23} \kappa^{-\frac23}. \label{p_LD}
\end{eqnarray}
%


From Eqs.~(\ref{p_MD}) and (\ref{p_LD}), there can be a third regime with $\pi_\text{LD}<\pi_\text{i}<\pi_\text{MD}$, where we expect the laser field to be too weak to drive the electron close to $\chi_\text{wave}^\text{MB}$ but at the same time strong enough for $\pi_\text{wave}$ to frequently change its sign. Hence, \re{FSSA} will \re{not be effective} here. We label this regime the \textit{deflection regime} (DR). It disappears with the increase of $\alpha$ at $\pi_\text{MD}=\pi_\text{LD}$, which occurs at $\alpha = \alpha^*\equiv a_0 \kappa/\rho$. An important conclusion is that the energy enhancement takes place regardless of the initial electron momentum beyond this threshold for $\alpha \geq \alpha^*$.




%

To now test these analytical estimates and fix the free parameters in Eqs.~(\ref{p_MD},\ref{p_LD}) we integrate Eq.~(\ref{exactdynamics}) numerically. We study a laser of amplitude $a_0=50$ and wavelength $\lambda = 1\,\mu$m, a \re{normalized} electron current density $\alpha=0.01$ and an electron injected into the plasma at the initial phase $\xi_0=\pi/2$. We begin by studying the MD regime, to which end we choose $\pi_\text{i}=100\gtrsim a_0$. In this regime, the laser field does not induce a sign flip of $\pi_\text{wave}$ and the electron will continuously approach the MB. We then study an exemplary electron trajectory (Fig.~\ref{fig2}a and~\ref{fig2}b), along which the rainbow color bar gives the dephasing $R$. We see that indeed the electron accumulates energy exclusively around the trajectory's turning points where $\chi_\text{wave}\to\chi_\text{wave}^\text{MB}$ ($R\to0$). In particular, at three turning points we study the electron's transverse position, relative to the laser field and energy gain $d\gamma/d\xi$ (Fig.~\ref{fig2}c). While at MD1 and MD2 $d\gamma/d\xi$ is antisymmetric in phase and the electron effectively absorbs energy from the laser field, at the turning point MD3 the net energy gain is small since it coincidentally occurs in phase with the laser field. In momentum space the electron's oscillatory behavior with discrete jumps to higher energy levels \re{is clearly visible} (Fig.~\ref{fig2}f). 

Next, we study the LD regime by choosing $\pi_\text{i}=5 \ll \pi_\text{LD}=13.5 \ll a_0$. In contrast to the MD regime the electron approaches the MB at every laser period (Fig.~\ref{fig2}d). As a result, the electron accumulates \re{more energy than achievable without the magnetic field} $\gamma\approx2000\approx4\gamma_0$ (Fig.~\ref{fig2}b). In momentum space we find the electron trajectory in the LD regime, in contrast to the MD regime, to jump to high energies within a single laser cycle (Fig.~\ref{fig2}f). 

To visualize the intermediate, non-accelerating DR regime we choose $\pi_\text{LD}<\pi_\text{i}=25<\pi_\text{MD}$. In this regime the electron does not approach the MB (black line in Fig.~\ref{fig2}a) and its minimum dephasing at the maximum transverse position is $R_\text{min}\approx14$. Consequently, the absorbed energy is always oscillating (Fig.~\ref{fig2}e) and the net energy gain is small (Fig.~\ref{fig2}b).

\begin{figure}[t]
\includegraphics[width=0.98\columnwidth]{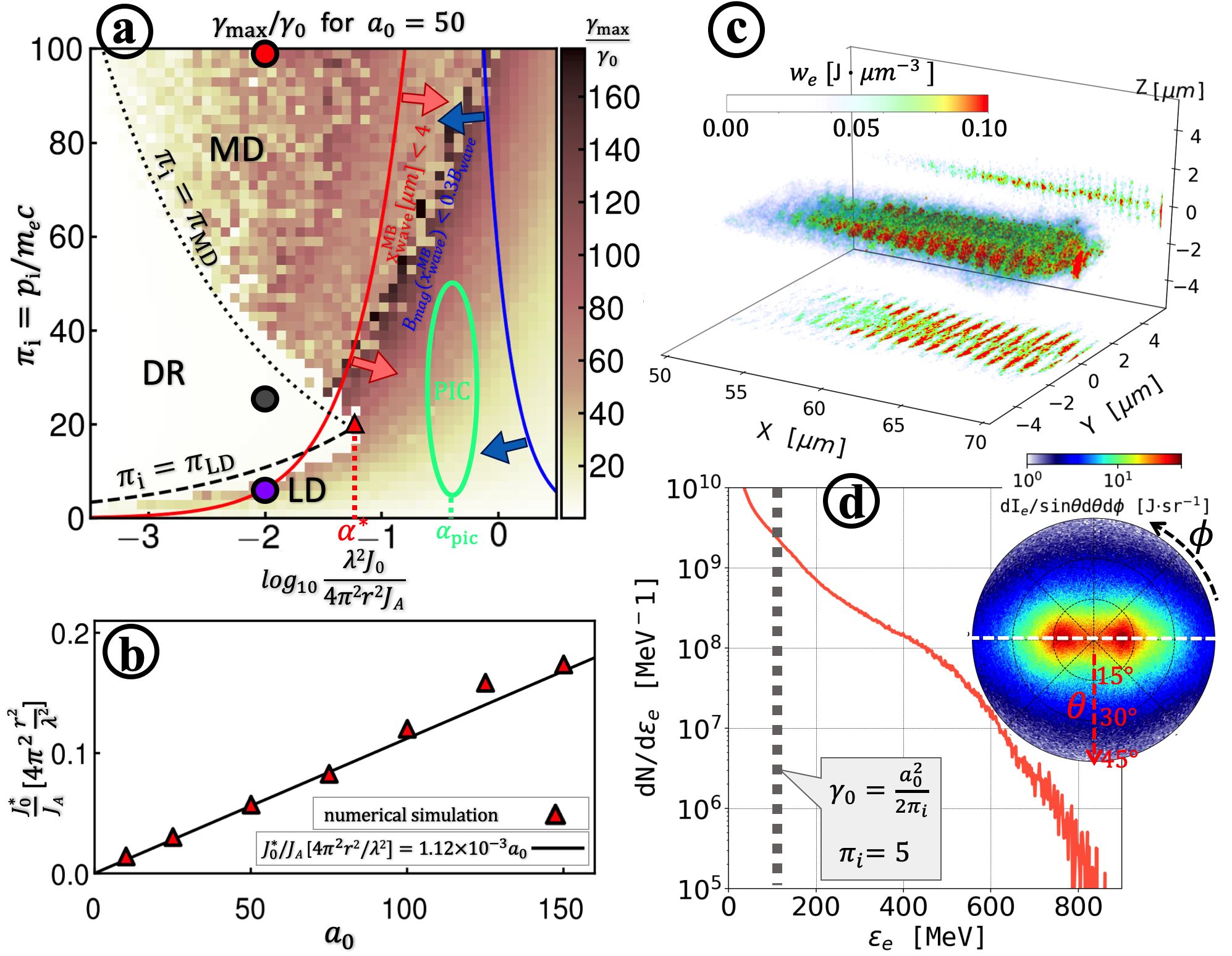}
\caption{(a) Numerical parameter scan of $\gamma_\text{max}/\gamma_0$ for $a_0=50$, where $\gamma_0=a_0^2/2\pi_\text{i}$. The red (MD), black (DR) and purple (LD) circles correspond to the three case in Fig.~\ref{fig2}. The black dotted and dashed curves represent the thresholds given by Eq.~(\ref{p_MD}) and~(\ref{p_LD}). The right (left) side of the solid red (blue) line denotes
the region for which $x_\text{wave}^\text{MB}<4\mu m$ ($B_\text{mag} (x_\text{wave}^\text{MB}) < 0.3 B_\text{wave}$). The red triangle indicates the threshold $\alpha^*$. The green ellipse marks the regime of the 3D simulations~\cite{Supplemental_Material} where $\alpha_{pic}\approx0.4$. (b) Comparison of numerical results (red triangles) to the analytical scaling of Eq.~(\ref{alpha_s}) (solid black line). (c) The distribution of electron energy density $w_e$ from the 3D PIC simulation. (d) The energy spectrum of the accelerated electrons in the 3D simulations, with the insert showing the angular distribution in the electron momentum space [see Supplemental Material for a detailed discussion].}
\label{fig3}
\end{figure}

To highlight the broad parameter range in which \re{FSSA} is feasible, we present a detailed numerical parameter scan of the electron's maximum energy gain $\gamma_\text{max}$ normalized to the energy gain without magnetic field $\gamma_0$ as a function of the initial momentum $\pi_\text{i}$ and the \re{normalized current} $\alpha$ (Fig.~\ref{fig3}a). Fitting the free parameters of the above derived scalings Eq.(\ref{p_MD}, \ref{p_LD}) we find $\rho\approx 125$ and $\kappa \approx 0.14$, yielding good agreement between the numerical data and the analytical curves. Furthermore, \re{with these parameters we find the current beyond which an electron is accelerated independently of $p_\text{i}$ to be 
\begin{equation}
J_0^* = 4\pi^2 \alpha^* (r/\lambda)^2 J_A \approx 0.76\, a_0 (r/\lambda)^2 \,\text{kA}. \label{alpha_s} 
\end{equation}
}


Two conditions must be satisfied for the discussed model to be applicable. The first condition is that the width of the laser beam $w_0$ must exceed the size of the region inside the magnetic boundary, i.e. $w_0 > x_\text{wave}^\text{MB}$. The second condition is that the plasma magnetic field inside the region set by the magnetic boundary must be significantly below the magnetic field of the laser, i.e. $B_\text{mag} (x_\text{wave}^\text{MB}) \ll B_\text{wave}$, for the laser to be able to drive this field. These conditions are not mutually exclusive, as illustrated in Fig.~\ref{fig3}a for $w_0 = 4\,\mu$m and $B_\text{mag} (x_\text{wave}^\text{MB}) < 0.3 B_\text{wave}$. The corresponding laser power is 1.7~PW. According to Eq.~(\ref{alpha_s}), the plasma current must exceed $J_0^* \approx 0.24 \mbox{ MA} \gg J_A$. This condition is satisfied in the 3D simulation presented in the Supplemental Material, with $J_0 \approx 1.7$~MA. Additionally, the condition $w_0 > x_\text{wave}^\text{MB}$ is satisfied for $\pi_i < a_0 = 50$. In agreement with our theory, the FSSA mechanism is realized in this regime, as evident from the electron spectrum in Fig.~\ref{fig3}d. The  localization  of the energetic  electrons is  confirmed  by  the  snapshot  of  the  electron  energy  density $w_e$ shown  in  Fig.~\ref{fig3}c. The angular  distribution  in  momentum  space (Fig.~\ref{fig3}d) confirms that the electron trajectories are flat. It must be pointed out that the current limit for a beam of relativistic electrons is $\gamma J_A$~\cite{auer1974self,dodin2006correction}, where $\gamma$ is the relativistic factor associated with the directed motion. In our simulation, there are electrons with $\gamma > 100$ needed to satisfy the criterion $J_0 < \gamma J_A$\cite{Supplemental_Material}. 


As confirmed by the simulations, the FSSA mechanism is realized due to the relativistically induced transparency~\cite{2012_Relativistic_transparency,stark2015relativistic,fernandez2017laser} in a structured target that mitigates instabilities~\cite{Stark2016_PRL,jiang2016microengineering,snyder2019relativistic,huang2017_hosing_PRE}. This implies that the electron density in the propagation region is $n_e \ll a_0 n_\text{cr}$, where $n_\text{cr} = m_e \pi c^2/\lambda^2 e^2 \approx 1.1 \times 10^{21}\, \text{cm}^{-3} \left(\lambda[\mu\text{m}]\right)^{-2}$ is the classical critical density. At high $a_0$, this density can be overcritical, $n_e > n_{cr}$. This is why the total charge of electrons with energies above 400 MeV exceeds 2 nC in our 3D simulation.

In conclusion, in this manuscript we \re{identified and characterized the novel, direct \textit{forward-sliding swing acceleration} (FSSA)} mechanism, \re{steered by} an intense laser propagating through a dense plasma \re{generating} an azimuthal magnetic field. In this field combination an electron accumulates kinetic energy exclusively close to the magnetic boundaries, beyond which the azimuthal magnetic field inhibits electron motion. We identified two regimes in which the electron can approach this boundary and demonstrated that for currents exceeding the threshold $J_0^*$ it is accelerated to high energies regardless of its initial momentum. This clarifies how electrons \re{absorb energy} from the studied, complex field structure and indicates that the FSSA mechanism can facilitate gamma-ray emission~\cite{nakamura2012high,ridgers2012dense,brady2012laser,ji2014_PRL,ji2014energy,Stark2016_PRL,huang2016_pre,gonoskov2017_prx,zgong_PPCF_2018}.


The work has been supported by the National Science Foundation (Grant No. 1632777), the National Basic Research Program of China (Grant No.2013CBA01502), and NSFC (Grant No. 11535001). The PIC code Epoch was in part funded by the UK EPSRC grants EP/G054950/1, EP/G056803/1, EP/G055165/1 and EP/ M022463/1


\section{Supplemental Material}

The model presented in our Letter is based on a configuration where an intense laser pulse propagates through a volume with a strong quasi-static azimuthal magnetic field. In what follows, we use fully relativistic kinetic three-dimensional (3D) and two-dimensional (2D) simulations to demonstrate that such a configuration is indeed achievable with the help of structured targets that can now be produced at leading target manufacturing facilities. 

The document consists of four sections. In Section I, we describe the setup of our particle-in-cell (PIC) simulations for laser-irradiated structured targets with a pre-filled channel. In Section II, we present the calculated structure of the electron current and quasi-static electric and magnetic fields in the channel. Our 3D simulations confirm that the laser pulse can propagate through the channel in a stable fashion while driving a Mega Ampere level current that sustains a strong quasi-static magnetic field. In Section III, we present electron tracking results that illustrate the laser-driven acceleration regime that is achieved due to the presence of the quasi-static magnetic field. We also show that the difference between the laser phase velocity and the speed of light is of secondary importance in the considered case. In Section IV, we summarize the key results of our kinetic simulations.



\begin{figure*}[htb]
\includegraphics[width=1.96\columnwidth]{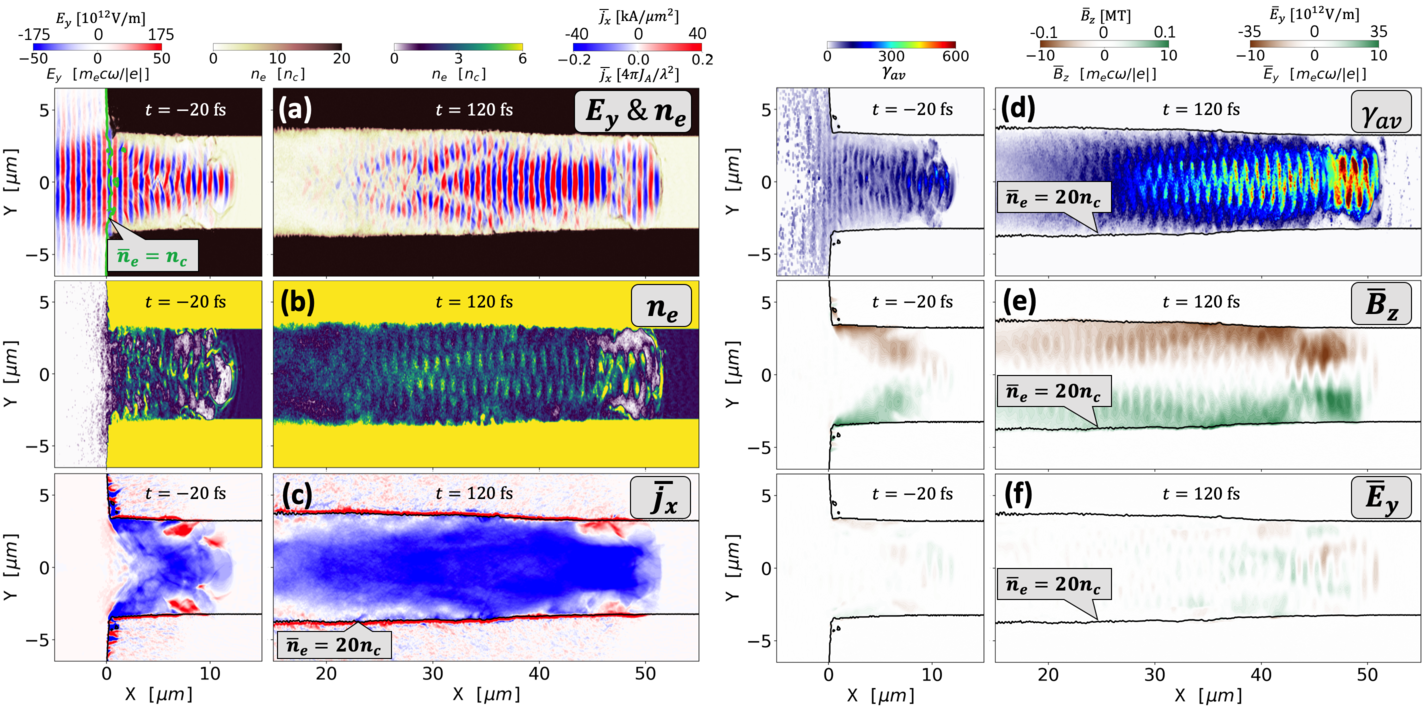}
\caption{3D PIC simulation snapshots in the $(x,y)$-plane at $z = 0$ and $t = 120$ fs. Smaller panels on the left show early snapshots taken at $t = - 20$ fs. (a) $E_y$ plotted on top of the electron density $n_e$, with the color saturated for $n_e > 20 n_c$. (b) Electron density $n_e$, with color saturated for $n_e > 6n_c$. (c) Time-averaged current density $j_x$. (d) Cell-averaged relativistic $\gamma$-factor $\gamma_{av}$. (e) and (f) Time-averaged magnetic and electric fields. The overline stands for time-averaging over 4 laser periods.}
\label{sup_field}
\end{figure*}


\bigskip \noindent \textbf{I. Particle-in-cell simulation setup --} In our 3D particle-in-cell (PIC) simulation, a laser pulse irradiates a uniform target with a pre-filled cylindrical channel. The central axis of the laser pulse is aligned with the central axis of the channel that is also the $x$-axis of the Cartesian system of coordinates used in our simulation (see Fig.~\ref{sup_field}a). The laser pulse is incident from the left and it is focused at the channel entrance located at $x = 0$ $\mu$m. 


In the absence of the target, the laser pulse has a Gaussian focal spot of 4.7 $\mu$m (full width at half-maximum for the laser intensity), with a peak intensity of $3.4\times 10^{21}$ watt/cm$^2$ and a normalized laser amplitude of $a_0 \approx 50$. It is linearly polarized with a wavelength of $\lambda = 1$ $\mu$m. The laser electric field in the focal plane is directed along the $y$-axis, while the magnetic field is directed along the $z$-axis. The time profile of the electric field at $x = y = z = 0$ $\mu$m is $|E_y| = E_0 \cos(\pi t/\tau)$ for $|t| < \tau/2$ and it is $|E_y|=0$ for $|t| > \tau/2$, where $E_0$ is the maximum field amplitude, $\tau = 160$ fs is the pulse duration, and $t=0$ fs is the time when the laser reaches its peak intensity in the focal plane. We choose this time profile in order to make our 3D simulation more manageable. The important physics takes place inside the laser pulse, so the laser pulse duration is more important in the context of our problem than the rise time.  

The target is initialized as fully ionized carbon, which is a good approximation for plastic targets. The electron density in the bulk is set at $n_e = 25 n_c$, where $n_c$ is the critical density that determines the electron density cutoff for linear electromagnetic waves with $\lambda = 1$ $\mu$m.
The target has a cylindrical channel with radius $R = 3.2$~$\mu$m. The initial electron density inside the channel is set at $n_e=1.5n_c$, so that it is opaque at laser amplitudes below $10^{18}$ watt/cm$^2$. We represent electrons by 20 macro-particles per cell, while the ions are represented by 10 macro-particles per cell. It is worth pointing out that structured targets with empty channels whose radius is several microns have already been used to conduct successful experiments~\cite{jiang2016microengineering,snyder2019relativistic}. It is feasible to fill these targets with a low density foam to achieve the considered configuration. Note that no ionization takes place during our simulation, which significantly reduces computational costs. This approach is justified by performing additional simulations with  field ionization. These  simulations  show  that  the leading edge of the considered laser pulse can fully ionize carbon well before the intensity reaches its peak value.

In order to achieve a significant electron energy gain in our simulation, we use a target that is 75 $\mu$m long. The electrons are accelerated as they move forward with the laser pulse, which is the reason why an extended target is required. The size of our simulation domain in the $(x,y,z)$-space is 80 $\mu$m $\times$ 24 $\mu$m $\times$ 24 $\mu$m and the cell size is (1/20) $\mu$m $\times$ (1/15) $\mu$m $\times$ (1/15) $\mu$m.

A well-resolved particle tracking procedure requires frequent outputs of particle data and electric field components. We achieve this by performing a 2D simulation with a setup similar to that used in 3D. In the $(x,y)$-space, the domain has the same size, but the cell size is smaller [(1/50) $\mu$m $\times$ (1/30) $\mu$m]. We use 40 macro-particles per cell for the electrons and  20 macro-particles per cell for ions. To track the electron dynamics, we output the simulation data ten times per laser period. The electrons are tracked over 400 fs. We have repeated the 2D simulation after doubling the spatial resolution. The energetic part of the resulting electron spectrum has remained unchanged, which confirms that our results are not sensitive to the resolution used for electron tracking.


\bigskip \noindent \textbf{II. Field and current structure in a relativistically transparent channel --} Figure~\ref{sup_field} presents results of the 3D PIC simulation described in Section I and illustrates three main features: laser propagation via relativistically induced transparency, generation of a strong longitudinal current, and the ability of the plasma to sustain an extreme volumetric quasi-static magnetic field. 

In our setup, the laser propagation through the classically overdense plasma, as seen in Fig.~\ref{sup_field}a, is achieved through the so-called relativistically induced transparency~\cite{2012_Relativistic_transparency, stark2015relativistic,fernandez2017laser}. The strong laser electric field energizes target electrons and increases the characteristic $\gamma$-factor, shown in Fig.~\ref{sup_field}d as a cell-averaged quantity $\gamma_{av}$. This changes the optical properties of the material by increasing the effective critical density to approximately $\gamma_{av} n_c$. For our set of target and laser parameters, the channel becomes very transparent, while the bulk of the target remains opaque, creating an optical wave-guide for the laser beam. As a result, the laser pulse maintains a relatively high amplitude over tens of microns, which is significantly longer than the Rayleigh range for this beam.

The channel inside the target is required for the stable laser pulse propagation. The target structure suppresses the hosing instability that would develop in a uniform target~\cite{huang2017_hosing_PRE}. Additional simulations for a uniform target with $n_e = 1.5n_c$ confirm that the laser pulse experiences a significant deviation from its original direction after propagating just tens of microns into the target.

The channel plays another important role -- it sustains a strong laser-driven electron current density shown in Fig.~\ref{sup_field}c. It is well-known that a tightly focused laser beam expels plasma electrons while propagating through an initially uniform target. A strong transverse electric field that arises in this case causes ions to expand radially as well. This quickly leads to a formation of an empty channel without an electron current. In our case, the transverse laser electric field continuously re-injects electrons from the channel walls and keeps the channel filled with electrons for hundreds of femtoseconds, as shown in Fig.~\ref{sup_field}b. The density in the channel remains well above $\overline{n}_e=n_c$, as evident from the position of the corresponding contour-line in Fig.~\ref{sup_field}a that remains effectively unchanged. The overline indicates time-averaging over four laser periods.

\begin{figure}[t]
\includegraphics[width=0.98\columnwidth]{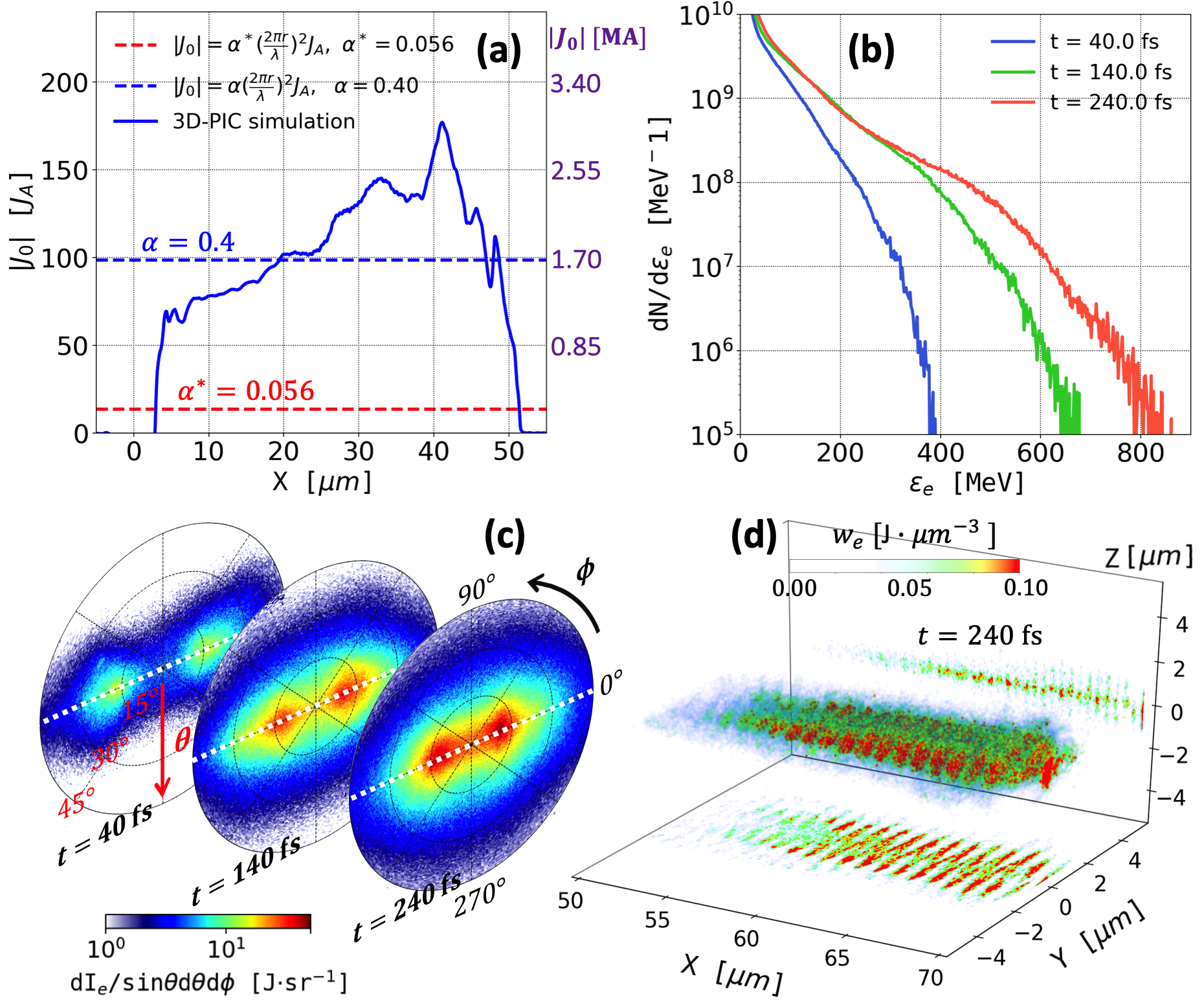}
\caption{Laser driven current (a) and energy spectra (b) and angular (c) and spatial (d) distributions of energetic electrons. The total current $J_0$ is the integral of $j_x$ over an area with $r < 2.5$~$\mu$m at $t=160$ fs. The red dashed line is the threshold current defined by Eq.(7) of the Letter. Panel (c) shows the electron energy $d I_e$ associated with a solid angle in momentum space (a detailed description is given in the text). Panel (d) shows the electron energy density $w_e$.}
\label{sup_en}
\end{figure}

The longitudinal electron current driven by the laser inside the channel exceeds the non-relativistic Alfv\'en current, $J_A \approx 17$ kA, by two orders of magnitude. The current density driven by the laser remains mostly uncompensated, because the return current is localized at the periphery of the laser beam~\cite{return_current_2,return_current_1}. In our case, most of the return current flows at the edge of the channel.  Figure~\ref{sup_en}a shows the total electron current, $J_0$, obtained by integrating the longitudinal electron current density $j_x$ over the area with $r < 2.5$~$\mu$m. This current exceeds $J_0 \approx 100 J_A \approx 1.7$~MA. It must be pointed out that the current limit for a beam of relativistic electrons is $\gamma J_A$~\cite{auer1974self,dodin2006correction}, where $\gamma$ is the relativistic factor associated with the directed motion. In our simulation, there are electrons with $\gamma > 100$ (see Fig.~\ref{sup_field}d) needed to satisfy the criterion $J_0 < \gamma J_A$.

In agreement with the assumption made in the Letter, the electron current density $j_x$ sustained by forward-moving electrons is nearly constant in the channel cross-section. It is well approximated by a constant value for $r < 2.5$~$\mu$m. We find that the normalized current density corresponding to $J_0 = 100 J_A$ (dashed blue line in Fig.~\ref{sup_en}a) is $\alpha \approx 0.4$. It is roughly seven times higher than the threshold value of $\alpha^* \approx 0.06$ for the FSSA that is given by Eq.~(7) of the Letter. The current corresponding to this threshold value, $J_0 \approx 14 J_A$, is represented by the dashed red line in Fig.~\ref{sup_en}a.

\begin{figure*}[t]
\includegraphics[width=1.95\columnwidth]{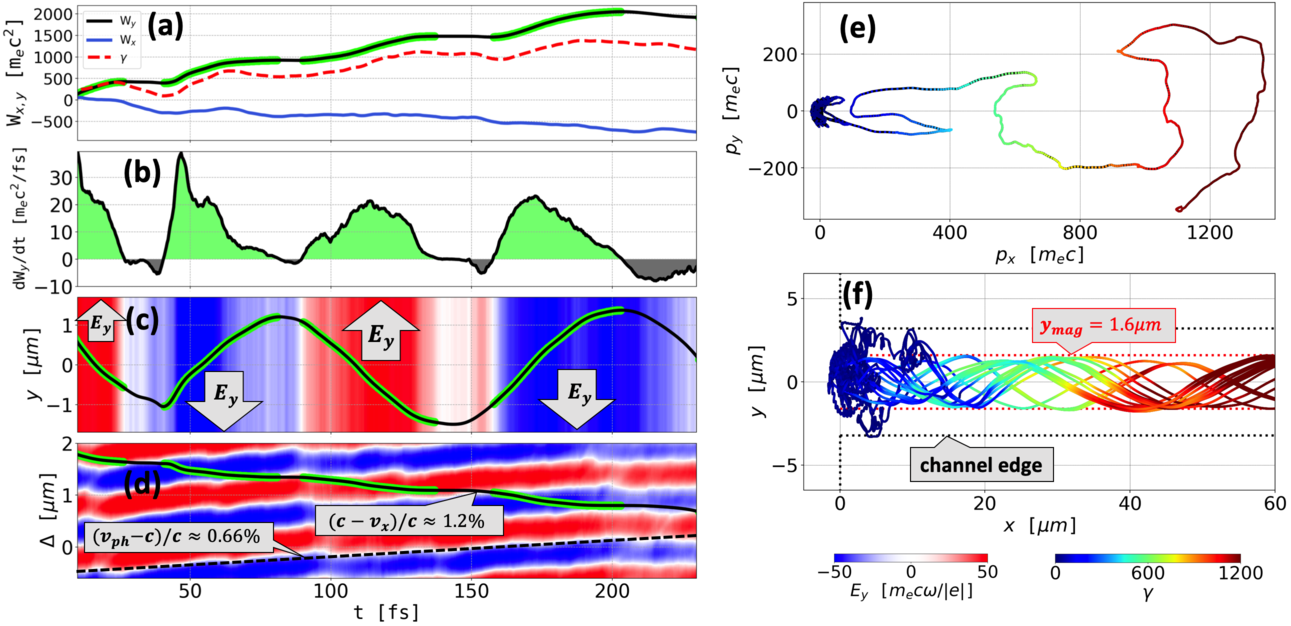}
\caption{Particle tracking from the 2D-PIC simulation. (a) Work by transverse and longitudinal electric fields. (b) The rate of the energy transfer by $E_y$ to the accelerated electron. (c) Transverse oscillations with respect to the wave-fronts of $E_y$. (d) Electron trajectory in a window moving with the speed of light. The background color is $E_y$ at $y = 0$ and the vertical coordinate is $\Delta=x-ct$. (e) The electron trajectory in the momentum space $(p_x, p_y)$. (f) Trajectories for 20 electrons whose energy exceeds 650 MeV at $t = 200$~fs. Note that the green color in panels (a)-(d) highlights the part of the trajectory where the electron is gaining energy from $E_y$.}
\label{sup_track}
\end{figure*}

The longitudinal electron current generates and sustains a strong quasi-static azimuthal magnetic field that can affect electron acceleration. The profile and strength of this field is shown in Fig.~\ref{sup_field}e. The color represents $\overline{B}_z$ in the $(x,y)$-plane at $z = 0$, where  the  overline again stands for time averaging. The field strength is consistent with a magneto-static calculation that assumes a cylindrically symmetric and uniform current density represented by the parameter $\alpha$:
\begin{eqnarray}
B \approx \alpha \frac{4\pi r}{\lambda} \frac{m_ec\omega}{|e|}. \label{B_field}
\end{eqnarray}
This expression yields $B \approx 12 m_e c \omega/|e| \approx 0.13$ MT at $r=2.5$ $\mu$m for $\alpha=0.4$, which is comparable to the result of the 3D simulation.

The laser beam also generates a transverse quasi-static electric field whose profile is shown in Fig.~\ref{sup_field}f. The color is the amplitude normalized the same way as the amplitude of the azimuthal magnetic field. We can conclude, by comparing Figs.~\ref{sup_field}e and \ref{sup_field}f, that the quasi-static electric field is relatively weak. The discussed electron injection into the channel prevents the laser beam from maintaining significant charge separation, which weakens the transverse electric field. At the same time, this is the mechanism that maintains the strong current and the resulting azimuthal magnetic field. 

The configuration where the quasi-static azimuthal magnetic field is much stronger than the quasi-static transverse electric field is rather generic at high laser intensities, i.e. $a_0 \gg 1$. The ion mobility provides an additional mechanism preventing a prolonged existence of strong transverse electric fields. For example, we found that the described configuration arises even in an initially empty channel~\cite{wang2019impact}. The charge of the injected electrons creates a radially inward electric field that drags ions from the channel walls, causing the channel interior to fill up. The characteristic time for this process scales as $a_0^{-1/2}$~\cite{wang2019impact}. It would take less than 50 fs for the channel to fill up and for the transverse electric field to be essentially eliminated for our parameters and an initially empty channel.

As shown in the Letter, the azimuthal magnetic field alone can significantly enhance the electron energy gain from the laser beam. This effect is confirmed by our 3D simulation. Snapshots of the electron spectra shown in Fig.~\ref{sup_en}b indicate that the electron energies continue to increase over time, with the maximum energy doubling in 200 fs. The energy increase is achieved in a scenario where the electron dynamics is dominated by the laser fields and the quasi-static magnetic field of the plasma, which underscores the importance of the developed theory for the FSSA mechanism.



\bigskip \noindent \textbf{III. Generation of energetic electrons --} In the previous section, we showed that the laser beam indeed creates conditions similar to the ones used to develop the FSSA model presented in the Letter. The electron spectrum, shown in Fig.~\ref{sup_en}b, confirms that the electron energies increase over time in this configuration, which is one of the signatures for the FSSA mechanism. In what follows, we take a close look at the spatial distribution of the energetic electrons and we also present detailed particle tracking that illustrates the main features of the FSSA mechanism.

The FSSA model involves several simplifications, one of which is the assumption that the electron trajectory is flat. The self-consistent 3D simulation allows us to examine and confirm the validity of this assumption. Figure~\ref{sup_en}c represents an angular distribution of energetic electrons in momentum space $(p_x, p_y, p_z)$, where $\phi = \arctan(p_z/p_y)$ is the azimuthal angle and $\theta=\arctan \left[ \left( p_y^2 + p_z^2 \right)^{1/2} / p_x \right]$ is the polar angle. In order to aid the visualization, we have plotted $dI_e/d \Omega$, where $dI_e$ is the electron energy associated with a solid angle $d \Omega = \sin \theta d \theta d \phi$ in momentum space. The momentum of an electron oscillating in the $(x,y)$-plane at $z = 0$ is shown with a white dotted line. For all three snapshots in Fig.~\ref{sup_en}c, most of the energy is concentrated near this line. These snapshots correspond to the energy spectra shown in Fig.~\ref{sup_en}b. We can thus conclude that the laser-accelerated electrons indeed tend to move along flat trajectories.

The localization of energetic electrons in the $(x,y)$-plane is further confirmed by the snapshot of the electron energy density $w_e$ shown in Fig.~\ref{sup_en}d. It is worth emphasizing that this is the polarization plane of the laser electric field. The transverse displacement of the energetic electrons in this plane is constrained to $|y| < 2$ $\mu$m, which is smaller than the channel radius, $R = 3.2$ $\mu$m. This confirms the transverse electron confinement by the azimuthal magnetic field in the region with a nearly uniform current density. The magnitude of the transverse displacement is in agreement with the estimate for the location of the magnetic boundary, $|y_{mag}| \approx (\lambda/2\pi)\sqrt{p_0/\alpha m_e c} \approx 1.8$ $\mu$m, where the initial transverse momentum is set at $p_0 \approx a_0 m_e c$ and $\alpha \approx 0.4$ is adopted {\color{black}from the blue dashed line} in Fig.~\ref{sup_en}a.

In order to obtain further details regarding the electron acceleration process, we perform particle tracking. Frequent data outputs for the electric field components, electron momenta, and electron locations are required. We achieve the desired time resolution by tracking the electrons in the 2D simulation whose parameters are described in Sec.~1. The 3D simulation has shown that the energetic electrons stay in the $(x,y)$-plane. That is why the 2D simulation with a laser electric field polarized in the plane of the simulation is a qualitatively reasonable approximation in terms of capturing the key physics, while it is also affordable in terms of post-processing. 

We have analyzed $5\%$ of the electrons that were randomly picked from the tail of the electron energy distribution ($\varepsilon_e >650$~MeV) at $t= 200$~fs. We found that the energy gain process and the electron trajectories have key similarities associated with the FSSA regime. For example, trajectories of 20 tracked electrons are shown in Fig.~\ref{sup_track}f. The dotted black lines mark the boundary of the bulk material. We find that the electrons are injected into the channel from its periphery close to the channel entrance. The electrons are being clearly confined in the transverse direction by the azimuthal magnetic field as they move along the channel and gain energy (the color indicates their $\gamma$-factor). Indeed, their transverse displacement, $|y| < 1.6$~$\mu$m, is significantly less than the transverse size of the channel, $|y| < 3.2$~$\mu$m.


The details of the acceleration process are shown in Figs.~\ref{sup_track}a - \ref{sup_track}e for a single electron from the tracked population. It is evident from the time evolution of the work done by transverse and longitudinal electric fields ($W_y$ and $W_x$ in Fig.~\ref{sup_track}a) that the electron energy is predominantly contributed by the transverse laser electric field $E_y$. Electron oscillations with respect to this field are shown in Fig.~\ref{sup_track}c, where the background color is the instantaneous electric field $E_y$ exerted on the electron. The azimuthal magnetic field changes the orientation of the transverse electron velocity and allows the velocity to remain anti-parallel to $E_y$ over extended segments of the electron trajectory (marked using the green outline). The electron gains energy while moving along these segments, as shown in Fig.~\ref{sup_track}b where the plotted quantity is $dW_y/dt=-(p_y/\gamma)E_y$. In agreement with our model, the electron energy is accumulated over multiple oscillations in the laser pulse. The trajectory of the considered electron in momentum space is shown in Fig.~\ref{sup_track}e. Its topology suggests that the electron is gaining energy in the laser-dominated (LD) regime of the FSSA.



Figure~\ref{sup_track}d shows the electron trajectory relative to the wave-fronts of the laser pulse in a window moving along the axis of the channel with the speed of light. Specifically, we define the coordinate in the moving window as $\Delta = x - ct$, whereas the background color is $E_y(y=0)$, which is the transverse laser electric field at the central axis. The moving window makes it easy to distinguish between superluminal and subluminal velocities. As one would expect, the wave-fronts have a positive slope, representing a superluminal phase velocity $v_{ph}$. We find that $v_{ph}-c \approx 6.6 \times 10^{-3} c$. This result quantifies the impact of the relativistically induced transparency on the laser propagation through the channel whose electron density is above $n_c$. The trajectory of the accelerated electron has a negative slope because the accelerated electron is moving slower than the speed of light. We find that $c - v_x \approx 1.2 \times 10^{-2} c$. In this case, the dephasing between the electron and the wave-fronts is primarily influenced by $v_x$ rather than by $v_{ph}$, since $c - v_x > v_{ph}-c$~\cite{robinson2015_vph}. One can then simplify the analysis of the electron dynamics by setting $v_{ph} \approx c$ and arrive to the FSSA model presented in the Letter. The validity of this approximation strongly depends on the level of the relativistically induced transparency and must be checked for a specific setup. 

\begin{figure}[t]
\includegraphics[width=0.98\columnwidth]{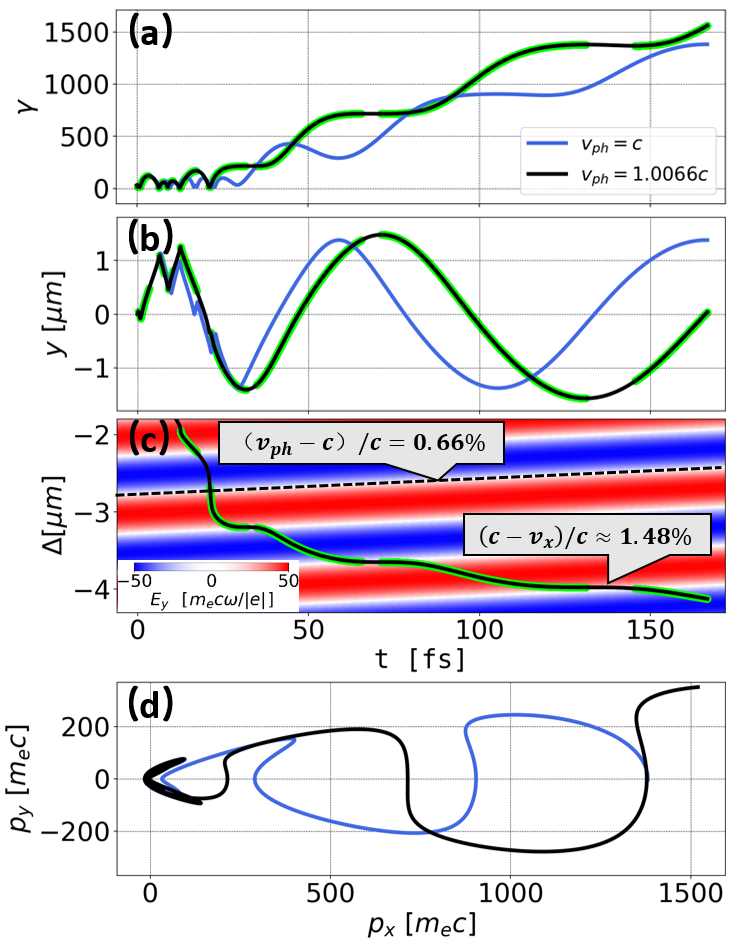}
\caption{Comparison of electron acceleration in a pulse with $v_{ph} = 1.0066c$ to electron acceleration at $v_{ph} = c$. (a) The time evolution of the $\gamma$-factor. (b) The transverse displacement $y$. (c) Electron trajectory in a window moving with the speed of light. The background color is $E_y$ and the vertical coordinate is $\Delta=x-ct$. (d) The electron trajectory in the momentum space $(p_x, p_y)$. Note that the green color in panels (a)-(c) highlights the part of the trajectory where the electron is gaining energy from $E_y$ at $v_{ph} = 1.0066c$. The initial transverse momentum is $p_0=30m_ec$ and the channel magnetic field has a strength corresponding to $\alpha=0.4$. }
\label{single_model}
\end{figure}

Figure~\ref{single_model} further confirms that the difference between the phase velocity and the speed of light has relatively little impact on electron acceleration for the parameters observed in our 2D PIC simulation. These plots are obtained by solving the equations of the FSSA model with $v_{ph} = c$ and with $v_{ph} = 1.0066 c$. The amplitude of the laser magnetic field is adjusted by a factor of $c/v_{ph}$ to be consistent with the Maxwell's equations. The initial transverse electron momentum is $p_0 = 30 m_ec$ in both cases. The value of $\alpha = 0.4$ is representative of the current density in the PIC simulations. In both cases, the electron $\gamma$-factor exceeds 1000, which is consistent with the particle tracking from the 2D PIC simulation.

\bigskip \noindent \textbf{IV. Summary --} We have demonstrated using self-consistent 3D and 2D kinetic simulations that a laser pulse can generate energetic electrons in an overdense plasma with a strong quasi-static magnetic field as a result of the FSSA mechanism. One advantage of this regime is that it allows the laser pulse to interact with a large number of electrons. The considered 160 fs laser pulse can generate 2.2 nC of electrons whose energies exceed 400 MeV (see $t = 240$ fs in Fig.~\ref{sup_en}b). Our simulations for a laser-irradiated structured target also validate the key assumptions and simplifications of the FSSA model by showing that (1) a strong current with $J_0 \gg J_A$ can be driven by the laser, with the current density being nearly constant in the region sampled by energetic electrons; (2) the laser propagation is stable and it is well-approximated by a plane wave with $v_{ph} \approx c$ due to the relativistically induced transparency; (3) the trajectories of energetic electrons remain flat and the magnetic field confines them radially well inside the laser beam.

\bibliography{aa.bib}

\end{document}